\documentclass[a4paper]{jpconf}
\usepackage{graphicx,amssymb}
\begin{document}
\title{Tests of the Electroweak Standard Model}

\author{Jens Erler}

\address{Departamento de F\'isica Te\'orica, Instituto de F\'isica, 
  Universidad Nacional Aut\'onoma de M\'exico, M\'exico D.F. 04510, M\'exico}

\ead{erler@fisica.unam.mx}

\begin{abstract}
Electroweak precision tests of the Standard Model of the fundamental interactions are reviewed
ranging from the lowest to the highest energy experiments. 
Results from global fits are presented with particular emphasis on the extraction of fundamental
parameters such as the Fermi constant, the strong coupling constant, the electroweak mixing angle, 
and the mass of the Higgs boson. 
Constraints on physics beyond the Standard Model are also discussed.
\end{abstract}

\section{Introduction}
The Standard Model (SM) of the electroweak (EW) interactions has been developed
mostly in the 1960s, where the gauge group $SU(2)_L \times U(1)_Y$ was 
suggested~\cite{Glashow:1961tr}, the Higgs mechanism for spontaneously broken
gauge theories developed, and the model for leptons constructed 
explicitly~\cite{Weinberg:1967tq}.  Subsequently, key predictions of the SM were
observed in the 1970s, including neutral currents and parity non-conservation 
in atoms and in deep-inelastic electron scattering (DIS)~\cite{Prescott:1978tm}.
The basic structure of the SM was established in the 1980s after
mutually consistent values of the weak mixing angle, $\sin^2 \theta_W$,
were determined from many different processes.
The 1990s saw the highly successful $Z$-factories, LEP and SLC, and 
the confirmation of the SM at the loop level.  It thus became clear that 
any new physics beyond the SM could at most be a perturbation.
The previous decade added precision measurements in the neutrino and quark sectors 
(including a 0.5\% measurement of the top quark mass~\cite{Aaltonen:2012ra}), 
as well as ultra-high precision determinations of the $W$-boson mass, 
$M_W$ (to $2\times 10^{-4}$)~\cite{TevatronElectroweakWorkingGroup:2012gb},
the anomalous magnetic moment of the muon~\cite{Bennett:2006fi},
and the Fermi constant, $G_F$~\cite{Webber:2010zf}.
These results suggest that the new physics must be separated by at least a little hierarchy 
from the EW scale unless one considers the possibility that a conspiracy is at work.
The current decade will elucidate the EW symmetry breaking sector at the LHC
and witness a new generation of experiments at the intensity frontier with sensitivities to
the multi-TeV scale and beyond. 

The next section reviews some recent developments in slightly more detail.
Interpretations of these and other results for the mass of the Higgs
boson, $M_H$, and new physics, are discussed, respectively, in the two sections thereafter.

\section{Recent Developments}
\subsection{Properties of charged leptons}

The MuLan Collaboration at the PSI in Switzerland~\cite{Webber:2010zf} has measured 
the $\mu$-lifetime to parts-per-million precision,
$\tau_\mu = 2.1969803(2.2) \times 10^{-6} \mbox{ s}$, which translates into a determination of
\begin{equation}
G_F = 1.1663787(6) \times 10^{-5} \mbox{ GeV}^{-2}.
\end{equation}
The Higgs vacuum expectation value is given by
$\langle 0|H|0\rangle= (\sqrt{2} G_F)^{-1/2} = 246.22$~GeV.
This measurement is so precise that even the error in the definition of the atomic mass unit (u) 
can shift $G_F$
(MuLan quotes $G_F = 1.1663788(7) \times 10^{-5} \mbox{ GeV}^{-2}$).
Moreover, it is so precise that the effect of the finite $M_W$ in the W-propagator 
is no longer negligible.
One may either choose to correct for it, {\em i.e.,} absorb it in $\Delta q$ defined through 
$\tau_\mu^{-1} \propto G_F^2 m_\mu^5 (1+ \Delta q)$, 
or else not to do so~\cite{vanRitbergen:1999fi}, 
{\em i.e.,} absorb it in $\Delta r$~\cite{Sirlin:1980nh} defined in terms of the accurately known 
fine structure constant, $\alpha$, and $Z$-boson mass, $M_Z$, 
\begin{equation}
  \sqrt{2} G_F M_W^2 \left( 1 - {M_W^2\over M_Z^2} \right) \equiv {\pi\alpha\over 1- \Delta r}\ .
\end{equation}
The latter convention is motivated by an effective Fermi theory point of view, 
and used by MuLan and since this year also by the PDG~\cite{PDG2012}.

What $\tau_\mu$ is to $G_F$ is the $\tau$-lifetime to the strong coupling constant, $\alpha_s$.
At least one low-energy $\alpha_s$-value is needed to promote the $Z$-width and related $Z$-pole
observables from a quantitative measurement in QCD to an EW SM test 
(or to constrain physics beyond the SM). 
Perturbative QCD has recently been extended to 4-loop order~\cite{Baikov:2008jh}, 
but there is a controversy whether the perturbative series should be truncated, {\em i.e.},
fixed order perturbation theory (FOPT) should be used~\cite{Beneke:2008ad}, or 
whether higher order terms from the running strong coupling in the complex plane should be 
re-summed in what is called contour-improved perturbation theory (CIPT)~\cite{Le Diberder:1992te}.
Unfortunately, FOPT and CIPT appear to converge to different values.
There are also non-perturbative contributions parametrized by condensate terms 
which can be constrained by experimentally determined spectral functions.
There are two different approaches~\cite{Davier:2008sk,Boito:2012cr}
which at present give very similar numerical results.
Using FOPT and the condensates from Ref.~\cite{Boito:2012cr},
\begin{equation}
\alpha_s [\tau] = 0.1193 \pm 0.0021, \hspace{72pt}
\alpha_s [Z] = 0.1197 \pm 0.0028,
\end{equation}
is found, where the latter determination from  the $Z$-pole is the only extraction of $\alpha_s$ with a very 
small theory uncertainty.  
The two values can be seen to agree perfectly.

The anomalous magnetic moment of the muon was measured to extreme precision,
\begin{equation}
\label{amu}
a_\mu \equiv {g_\mu - 2\over 2} = (1165920.80 \pm 0.63) \times10^{-9},
\end{equation}
by the BNL--E821 Collaboration~\cite{Bennett:2006fi}.
The prediction,
$a_\mu = (1165918.41 \pm 0.48) \times 10^{-9}$,
from the SM includes $e^+ e^-$ as well as $\tau$-decay data in the dispersion integral
needed to constrain the two- and three-loop vacuum polarization contributions and
differs by $3.0~\sigma$.
The data based on $\tau$-decays requires an isospin rotation and a corresponding
correction to account for isospin violating effects and suggests a smaller ($2.4~\sigma$)
discrepancy, while the $e^+ e^-$-based data sets (from annihilation and radiative returns)
by themselves would imply a $3.6~\sigma$ conflict.  
Indeed, there is a $2.3~\sigma$ discrepancy between the experimental branching ratio,
$B(\tau^- \to \nu \pi^0 \pi^-)$, and its SM prediction using the $e^+ e^-$ data~\cite{Davier:2010nc}.
In view of this, it is tempting to ignore the $\tau$-decay data and blame the difference
to the $e^+ e^-$ data on unaccounted for isospin violating effects. 
However, there is also a $1.9~\sigma$ experimental conflict between KLOE and BaBar 
(both using the radiative return method~\cite{Arbuzov:1998te})
the latter not being inconsistent with the $\tau$-data.

The above results include an additional uncertainty from hadronic 3-loop light-by-light scattering diagrams
which contribute, $\Delta a_\mu(\gamma\times \gamma) = (1.05 \pm 0.26)\times10^{-9}$~\cite{Prades:2009tw}.
This is consistent with the 95\% CL upper bound, $\Delta a_\mu(\gamma\times \gamma) < 1.59 \times10^{-9}$, 
found in Ref.~\cite{Erler:2006vu}.
One may point out that if the three dominant errors from experiment ($6.3\times10^{-10}$),
hadronic vacuum polarization ($3.6\times10^{-10}$) and light-by-light scattering ($2.6\times10^{-10}$)
can be pushed below $3\times10^{-10}$, 
then a $5~\sigma$ discovery would be established (if the central value persists).  
The nominal $\Delta a_\mu(\gamma\times \gamma)$ error is already there, but it is also the hardest to defend.

As for the question whether the deviation in $a_\mu$ may arise from physics beyond the SM
(especially supersymmetry~\cite{Ellis:1982by}), my personal take is that  I am less concerned 
about these hadronic issues than the absence of convincing new physics hints at the Tevatron or the LHC.

\begin{figure}
\begin{center}
\includegraphics[height=.48\textheight]{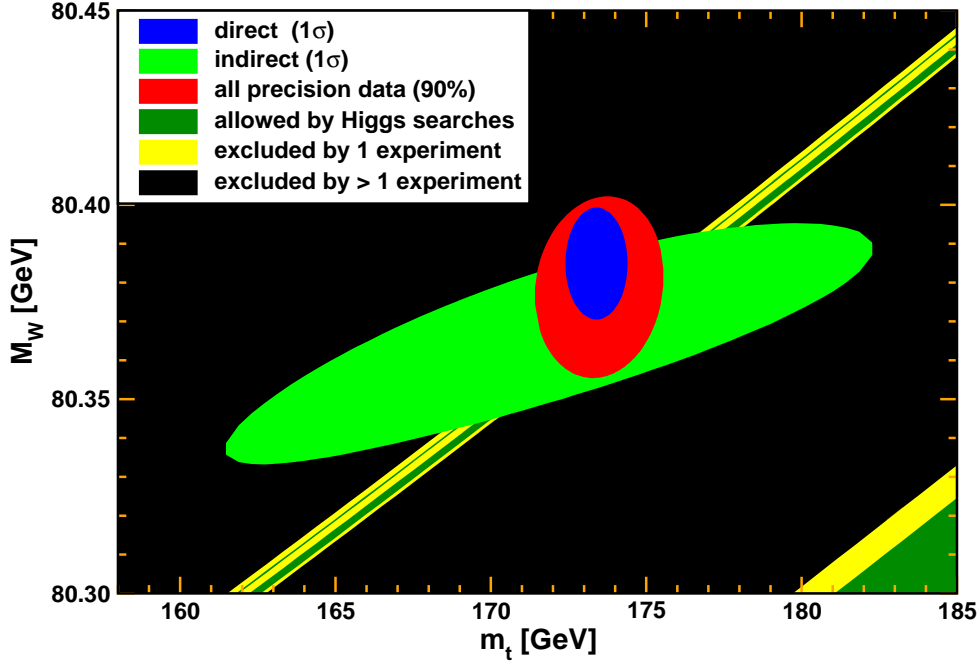}
\end{center}
\caption{\label{mwmt}$1~\sigma$ region in $M_W$ {\em vs.}\ $m_t$ for the direct and indirect 
EW precision data, and the 90\% CL region ($\Delta \chi^2 = 4.605$) allowed by both data sets. 
The SM predictions are also indicated, where the dark (green) bands correspond to $M_H$ 
which are currently allowed at the 95\% CL.
The bright (yellow) bands are excluded by one experiment and
the remaining (black) regions are ruled out by more than one experiment.}
\end{figure}

\subsection{$\sin^2\theta_W^{\rm on-shell}$}
The most precise derived and purely EW precision observable is no longer the $Z$-pole combination of 
$\sin^2\theta_W$, but rather $M_W = 80.387 \pm 0.016$ GeV from the CDF and D\O\ Collaborations
at the Tevatron~\cite{TevatronElectroweakWorkingGroup:2012gb} which is dominated by 
a $\pm19$~MeV determination by CDF using only 2.2~fb$^{-1}$ of their data.
Together with the LEP~2 combination~\cite{Alcaraz:2006mx}, $M_W = 80.376 \pm 0.033$~GeV, one 
obtains for the on-shell definition of $\sin^2\theta_W$,
\begin{equation}
\sin^2\theta_W^{\rm on-shell} \equiv 1 - {M_W^2\over M_Z^2} = 0.22290 \pm 0.00028,
\end{equation}
from which and can extract $M_H = 96^{+29}_{-25}$~GeV.  
The prospects for the full 10~fb$^{-1}$ dataset are a $\pm 13$~MeV $M_W$ determination
from CDF alone, even when no reduction of the parton distribution function (PDF) ($\pm 10$~MeV) 
and QED ($\pm 4$~MeV) uncertainties is assumed.
In the most optimistic scenario, CDF could shrink the error to $\pm 10$~MeV,
which is to be compared with the $\pm 6$~MeV accuracy expected from a threshold scan
at a future International Linear Collider.
The current direct and indirect determinations of $M_W$ and the top quark mass, $m_t$, 
are compared in Figure~\ref{mwmt}.

\begin{figure}
\begin{center}
\includegraphics[height=.48\textheight]{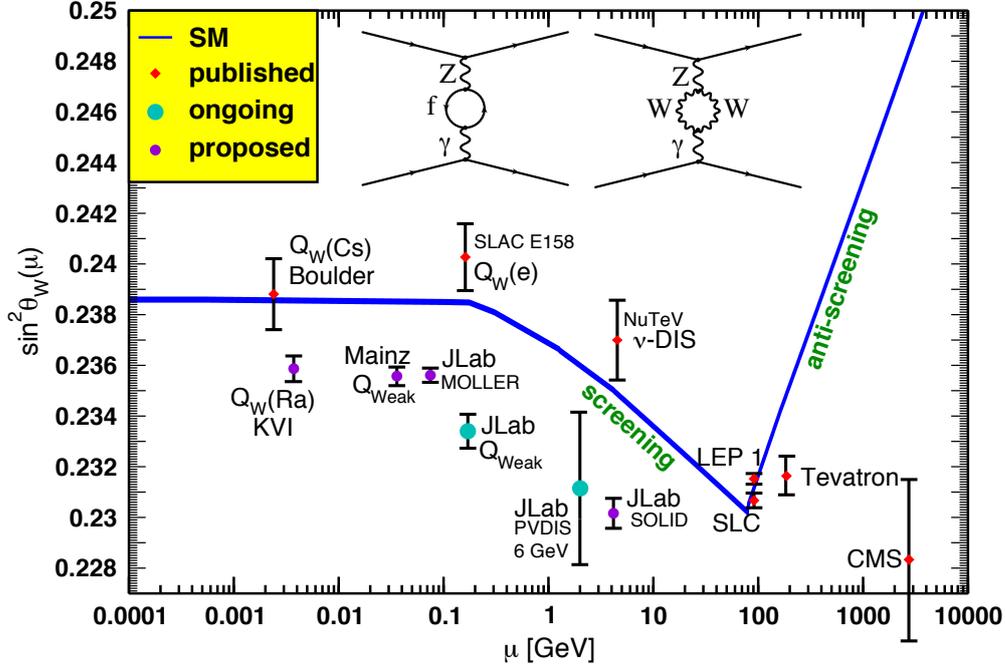}
\end{center}
\caption{\label{sin2theta}Current and future measurements of the running weak mixing angle. 
The uncertainty in the prediction is small except possibly in the hadronic transition region roughly
between 0.1 and 2~GeV~\cite{Erler:2004in}.  
The relevant $Q^2$ of the Tevatron and CMS values make them effectively additional $Z$-pole
measurements, but for clarity they have been shifted horizontally to the right.}
\end{figure}

Values of $\sin^2\theta_W^{\rm on-shell}$ are also often quoted from $R_\nu$, $R_{\bar\nu}$ or $R^-$,
which are combinations of $\nu$-DIS and $\bar\nu$-DIS neutral-current (NC) and charged-curent (CC) cross sections,
\begin{equation}\label{nudis}
R_\nu \equiv {\sigma_{\nu N}^{\rm NC} \over \sigma_{\nu N}^{\rm CC}}\ , \hspace{36pt}
R_{\bar\nu} \equiv {\sigma_{\bar\nu N}^{\rm NC} \over \sigma_{\bar\nu N}^{\rm CC}}\ , \hspace{36pt}
R^- \equiv \frac{\sigma_{\nu N}^{\rm NC} - \sigma_{\bar{\nu} N}^{\rm NC}}{\sigma_{\nu N}^{\rm CC} - \sigma_{\bar{\nu} N}^{\rm CC}} \approx {1\over 2} - \sin^2\theta_W.
\end{equation}
It should be cautioned, however, that $\sin^2\theta_W^{\rm on-shell}$ is affected very differently 
by variations of $M_H$ and by physics beyond the SM than the quantities defined in Eq.~(\ref{nudis}).
The most precise determination has been obtained by the NuTeV Collaboration~\cite{Zeller:2001hh}
at Fermilab,
\begin{equation}\label{nutev}
\sin^2\theta_W^{\rm on-shell} = 0.2277 \pm 0.0016,
\end{equation}
which is $3.0~\sigma$ above the SM prediction, 
$\sin^2\theta_W^{\rm on-shell} = 0.22296 \pm 0.00028$.
The deviation sits in the left-handed effective quark coupling, $g_L^2$, which is $2.7~\sigma$ off.
Various SM effects have been suggested, such as 
an asymmetric strange sea,
isospin violation (both, from QED splitting effects~\cite{Gluck:2005xh}
and through the PDFs~\cite{Sather:1991je}),
or nuclear effects ({\em e.g.}, the so-called isovector EMC effect~\cite{Cloet:2009qs}), as well as
QED~\cite{Arbuzov:2004zr}, QCD~\cite{Dobrescu:2003ta} and EW~\cite{Diener:2005me} radiative corrections.
We stress that the precise impact of these effects need to be evaluated carefully by 
the collaboration with a new and self-consistent set of PDFs, including new radiative 
corrections, while simultaneously allowing isospin breaking and asymmetric strange seas. 
This effort is currently on its way.

\begin{figure}
\begin{center}
\includegraphics[height=.48\textheight]{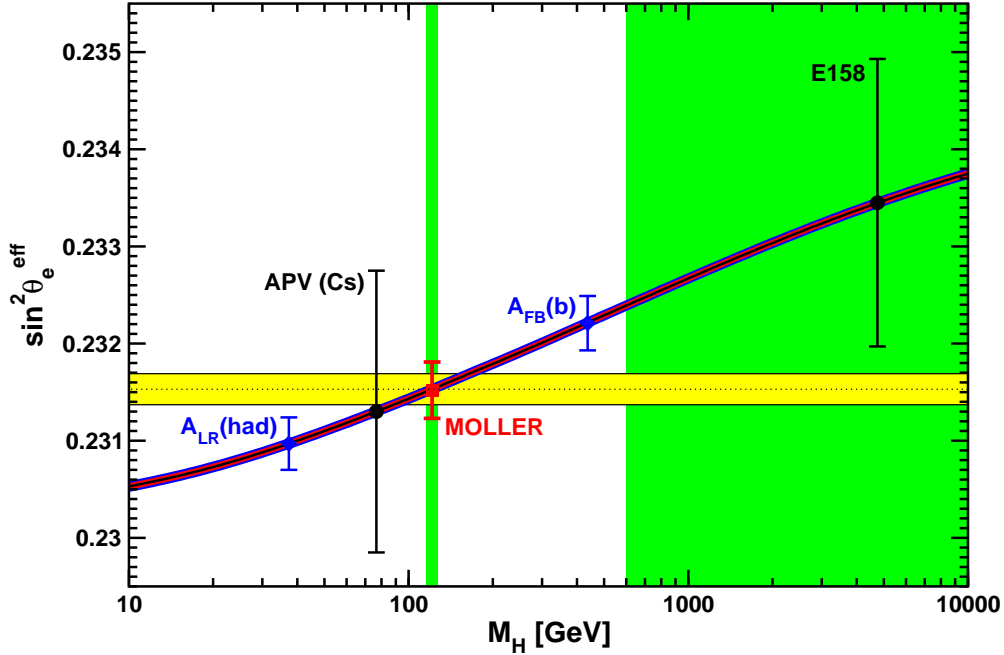}
\end{center}
\caption{Implication of $\sin^2\theta_W$ measurements for $M_H$. 
Shown are the most precise determinations from LEP~1 and the SLC, and the extractions from APV
and from current (E158) and future (MOLLER) polarized M\o ller scattering.
Also indicated are the non-excluded intervals from direct Higgs searches.
\label{KK}}
\end{figure}

\subsection{Parity-violating electron scattering}
High precision measurements in the EW sector are also possible at the intensity frontier,
when QED and QCD effects are filtered out by using parity-violating observables.

The JLab Qweak detector~\cite{Armstrong:2012ps} at the 6~GeV CEBAF was dedicated to a measurement 
of the weak charge of the proton, $Q_W^p \propto 1-4 \sin^2\theta_W$, to 4\% precision 
in elastic polarized $e^- p$ scattering at $Q^2 = 0.026$~GeV$^2$.
Data taking is complete and the analysis is in progress.
$Q_W^p$ is similar to the weak charges of heavy nuclei measured in atomic parity violation (APV) 
but at a different kinematics. 
This circumstance results in a re-enhancement of the $\gamma-Z$ box 
contribution~\cite{Gorchtein:2008px,Blunden:2011rd} introducing an extra theory uncertainty. 

The $\gamma-Z$ box is less of an issue at lower $Q^2$ which is one of the reasons why a similar 
experiment is also planned at a future facility (MESA) in Mainz at $Q^2 = 0.0022$~GeV$^2$. 
The projected uncertainties for $Q_W^p$ and the extracted $\sin^2\theta_W$ are 2.1\% and 
$\pm 0.00037$, respectively.

MOLLER~\cite{Mammei:2012ph} is an ultra-high precision measurement of $\sin^2\theta_W$ 
in polarized M\o ller scattering at the 12~GeV upgraded CEBAF~\cite{Dudek:2012vr}.
It aims at a factor of 5 improvement over a similar experiment at SLAC 
by the E158 Collaboration~\cite{Anthony:2005pm}, and would be one of the worlds most precise 
determinations of $\sin^2\theta_W$ and the most accurate at low energies. 

PVDIS was a deep-inelastic polarized $e^-$ scattering experiment using the 6~GeV CEBAF
and is currently in the analysis phase~\cite{Subedi:2011zz}. Together with SOLID (at 12~GeV) 
an array of kinematics points will be measured to test strong, EW, and new physics.
Figure~\ref{sin2theta} summarizes these and other current and future (projected) determinations
of $\sin^2\theta_W$ as a function of energy scale $\mu$.

\begin{figure}
\begin{center}
\includegraphics[height=.48\textheight]{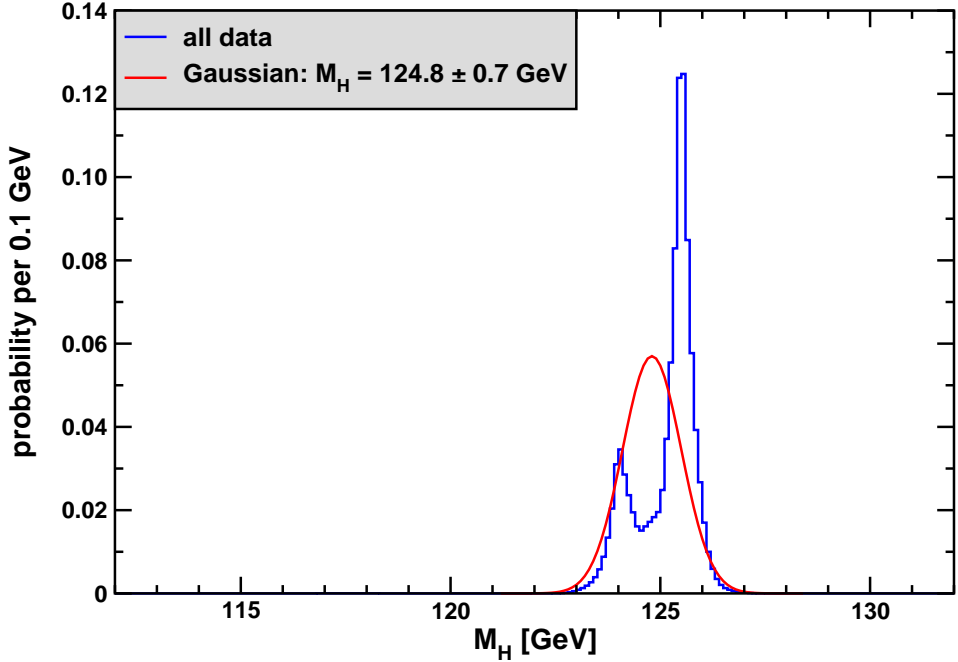}
\end{center}
\caption{The histogram shows the normalized probability distribution of $M_H$.   
The bell shaped curve is a reference Gaussian density defined to contain the same probability 
as the histogram over the region of bins which are higher than the tail bins. 
The significance of this region corresponds to $3.4~\sigma$.
\label{mh}}
\end{figure}

\section{SM Interpretation: $M_H$}
The various $\sin^2\theta_W$ measurements discussed above can be used to constrain $M_H$ 
and compare it with the results obtained at the LHC.  
It is important to recall that the most precise determinations at LEP~1 
(from the forward-backward cross-section asymmetry of $Z$-bosons 
decaying into $b\bar b$ pairs, $A_{FB}(b)$) and at the SLC 
(from the polarization asymmetry for hadronic final states, $A_{LR}({\rm had})$), 
both of which being mostly sensitive to the initial state (electron) coupling, 
are discrepant by three standard deviations. 
Their average, on the other hand, corresponds to values of $M_H$ that are in perfect agreement
with the Higgs boson candidates seen by the ATLAS~\cite{:2012an}
and CMS~\cite{Chatrchyan:2012tx} Collaborations at the LHC. 
This is illustrated in Figure~\ref{KK} together with the low-energy determinations from 
E158~\cite{Anthony:2005pm} and APV which is dominated by the experiment in 
Cs~\cite{Wood:1997zq} and makes use of the atomic theory calculation\footnote{After the conference
had adjourned there appeared an update of the atomic structure calculation~\cite{Dzuba:2012kx} 
finding significant corrections to formally subleading terms.
Taking this into account moves the extracted Cs weak charge $1.5~\sigma$ below the SM 
prediction, which then favors lower values of $M_H$.} 
of Ref.~\cite{Porsev:2010de}.

Estimating the significance of the LHC data~\cite{:2012an,Chatrchyan:2012tx} by themselves 
requires a ``look elsewhere effect correction" which is, however, poorly defined.  
It can be avoided when they are combined with the Higgs search results 
from LEP~2~\cite{Barate:2003sz} and the Tevatron~\cite{TEVNPH:2012ab} as
well as with the EW precision data~\cite{Erler:2012uu}, 
the latter providing a normalizable probability distribution\footnote{After the conference had 
adjourned there appeared updates of the LHC~\cite{Aad:2012gk,Chatrchyan:2012gu} and 
Tevatron~\cite{CDFandD0:2012zzl} Higgs search results including the announcement 
of the observation of a new boson consistent with the SM Higgs.  
For an update of Figure~\ref{mh} reflecting the new data, see Ref.~\cite{Erler:2012uu}.} 
shown in Figure~\ref{mh}.
This requires the validity of the SM which used to be a very strong assumption in the past.
But with the absence of clear new physics signals at the energy frontier 
this can now be seen as a reasonable approximation.

\begin{figure}
\begin{center}
\includegraphics[height=.48\textheight]{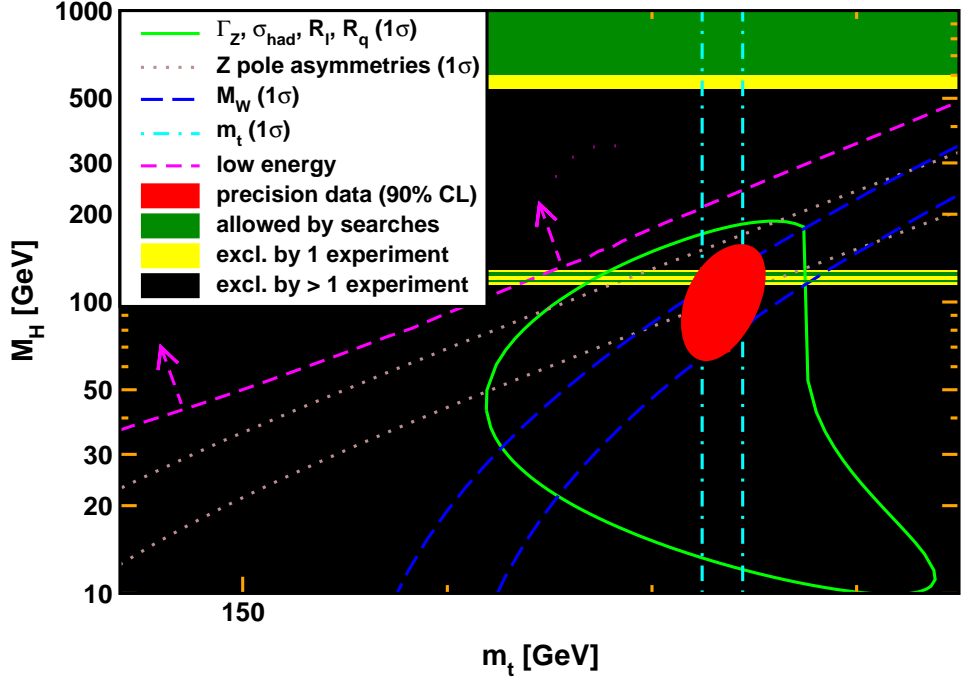}
\end{center}
\caption{$1~\sigma$ contours of $M_H$ as a function of the top quark
pole mass for various inputs, and the 90\% CL region allowed by all data. 
The color codes for the bands are as in Figure~\ref{mwmt}.
\label{mhmt}}
\end{figure}

\begin{figure}
\begin{center}
\includegraphics[height=.48\textheight]{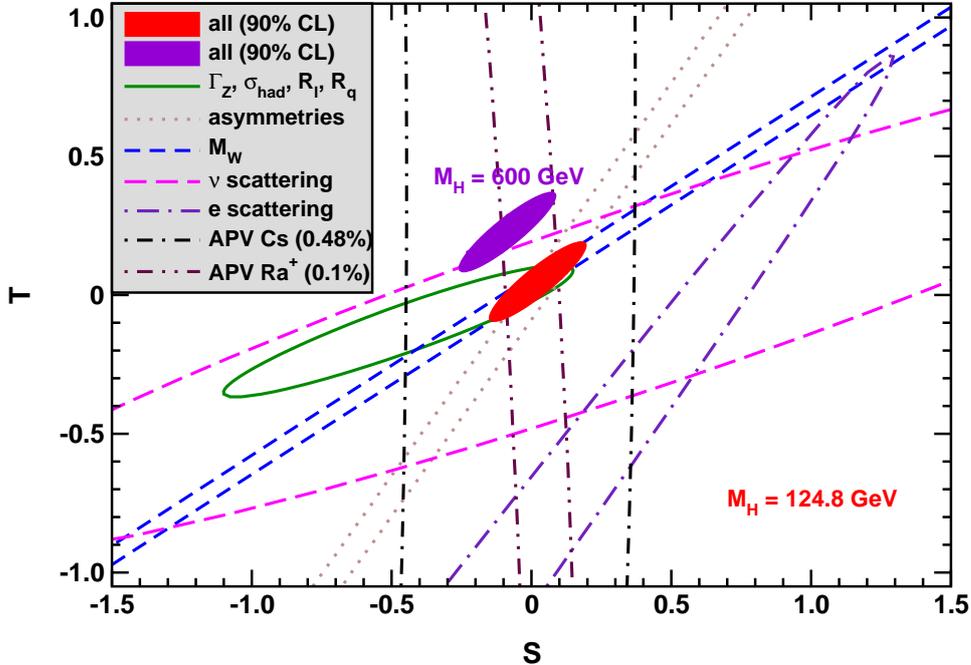}
\end{center}
\caption{1~$\sigma$ contours in $S$ and $T$ from various inputs assuming 
$M_H = 124.8$~GeV except for the upper (violet) one for all data  which is for $M_H = 600$~GeV.
The contour labeled APV Ra$^+$ refers to a {\em future\/} experiment on a single trapped Ra ion
which is in preparation at the KVI in Groningen~\cite{Jungmann:2008}.
The atomic structure of Ra$^+$ is alkali-like so that the atomic theory parallels that of Cs, 
but due to its greater neutron excess, Ra constrains a linear combination of $S$ and $T$ which is
different from Cs and quite orthogonal to the $M_W$ and $\sin^2\theta_W$ contours.
\label{ST}}
\end{figure}

As for the global EW fit without the collider events, I currently find,
\begin{equation}\label{eqmh}
M_H = 102^{+24}_{-20} \mbox{ GeV}.
\end{equation}
To extract values of $M_H$ from the EW precision data, as for example in Eq.~(\ref{eqmh}),
it is important to know the top quark mass, $m_t$, to very high precision. 
The various measurements from the Tevatron~\cite{Lancaster:2011wr} and 
the LHC~\cite{LHCmt} (strongly dominated by the CMS $\mu +$jets channel) combine to,
\begin{equation}
m_t = 173.21 \pm 0.51_{\rm uncorr} \pm 0.75_{\rm corr} \pm 0.5_{\rm theo} \mbox{ GeV} 
= 173.2 \pm 1.0 \mbox{ GeV},
\end{equation}
where I assumed that the Tevatron systematic error is common to both colliders, 
and where I have added a theory uncertainty from the relation~\cite{Chetyrkin:1999qi} 
between the top quark pole mass and the $\overline{\rm MS}$-mass definitions 
(the size of the three-loop term).  
It is moreover assumed that the kinematic mass extracted from the collider events corresponds 
within this uncertainty to the pole mass (see also Ref.~\cite{Skands:2007zg}).
Further improvements will eventually require to extract definitions of $m_t$ which are
easier to interpret theoretically. 
For example, by combining the effective theories for soft collinear (SCET) 
and heavy quarks (HQET) one can form the jet mass of Ref.~\cite{Fleming:2007qr}.
Alternatively, one may extract the $\overline{\rm MS}$-mass,
$\bar m_t(\bar m_t) = 160.0 \pm 3.3$~GeV,
which is the one actually entering the global EW analysis, 
directly from the $t\bar t$ cross-section~\cite{Langenfeld:2009wd}
(this corresponds to a pole mass of $m_t = 169.6 \pm 3.5$~GeV 
and would yield $M_H = 81^{+32}_{-24}$~GeV instead of Eq.~(\ref{eqmh})).
Constraints on $M_H$ as a function of $m_t$ are shown in Figure~\ref{mhmt} for various data sets.

\section{New Physics Interpretations}
\subsection{Oblique new physics}

The EW precision tests also set strong constrains on models of new physics.
{\em E.g.}, an extra generation or anti-generation of fermions is severely constrained~\cite{Erler:2010sk},
and if the Higgs hints are real, these are ruled out\footnote{After the conference
had adjourned there appeared a new analysis~\cite{Eberhardt:2012gv} raising the exclusion to 
the 5.3~$\sigma$ CL.} at the 99.6\% CL~\cite{Kuflik:2012ai}.
This leaves us with basically three scenarios, all of which in need of some tuning and faith
(the mass spectra are generally quite similar):
(i) One ignores the collider bumps (or assigns them to something else)
and assumes $M_H \lesssim 120$~GeV (see e.g., Ref.~\cite{Dighe:2012dz});
(ii) one assumes instead $M_H \gtrsim 450$~GeV~\cite{Buchkremer:2012yy};
(iii) or one accepts $M_H \approx 125$~GeV and introduces new physics 
beyond a fourth generation, such as an extra Higgs doublet~\cite{Bellantoni:2012ag}.

More generally, whenever the new physics is rather heavy and mostly affects 
the gauge boson self-energies, one can parametrize it in terms of the oblique 
parameters $S$ and $T$~\cite{Peskin:1991sw}
(a third parameter, $U$, is usually small).
The constraints on $S$ and $T$ from various data sets are shown in Figure~\ref{ST}
(where $U = 0$ is assumed).  
The three-parameter fit result is shown in Table~\ref{stu}.

\begin{figure}
\begin{center}
\includegraphics[height=.48\textheight]{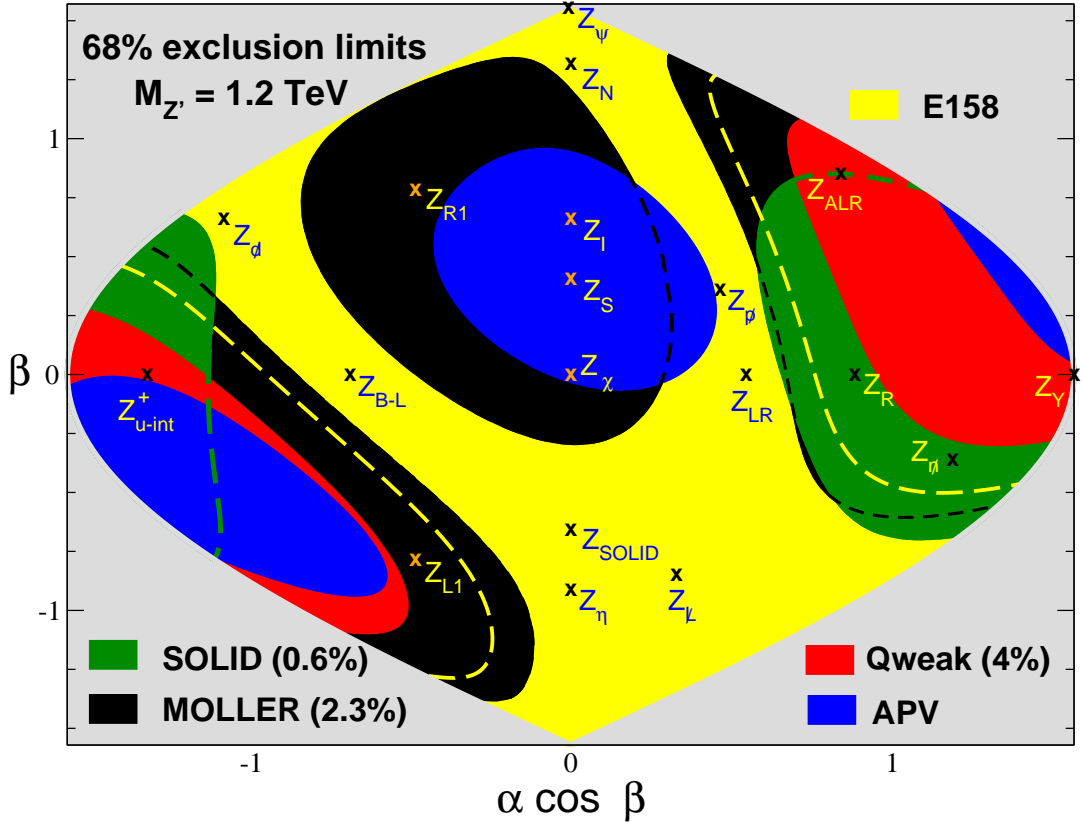}
\end{center}
\caption{68\% exclusion constraints on the $E_6$ parameters $\alpha$ and $\beta$ 
for $M_{Z^\prime} = 1.2$~TeV from various actual and hypothetical low energy measurements
(for future measurements it is assumed that the central values will coincide with the SM).
\label{Zprimes}}
\end{figure}

\subsection{Extra $Z$ bosons}

Among the best motivated kinds of physics beyond the SM are additional neutral $Z^\prime$ 
bosons~\cite{Langacker:2008yv} which are {\em not\/} well described by oblique parameters.
They easily appear in top-down scenarios like Grand Unified Theories or superstring constructions. 
In fact, it often requires extra assumptions if one wants to {\em avoid\/} an additional $U(1)^\prime$ gauge symmetry 
or decouple the associated $Z^\prime$ from observation. 
This is even more true in bottom-up approaches where $U(1)^\prime$ symmetries are a standard tool 
to alleviate problems in models of dynamical symmetry breaking, supersymmetry, large 
or warped extra dimensions, little Higgs, {\em etc.} 
And as all these models are linked to electroweak symmetry breaking, the $Z^\prime$ mass, $M_{Z^\prime}$, 
should be in the TeV region, providing a rationale why they might be accessible at current or near future experiments. 

\begin{table}
\caption{\label{stu} Result (including correlations) 
of the global fit to $S$, $T$, and $U$ for $M_H = 125$~GeV.} 
\begin{center} 
\begin{tabular}{ccrrr}
\br
parameter & fit result & \multicolumn{3}{c}{correlations} \\  
\mr
$S$ & $0.00 \pm 0.10$ &1.00 & 0.89 & $-0.55$ \\
$T$ & $0.02 \pm 0.11$ & 0.89 & 1.00 & $-0.80$ \\
$U$ & $0.04 \pm 0.09$ & $-0.55$ & $-0.80$ & 1.00 \\
\br
\end{tabular}
\end{center}
\end{table}

$Z^\prime$ discovery would most likely occur as an $s$-channel resonance at a collider, 
but interference with the photon or the standard $Z$ provides leverage also at lower energies. 
Once discovered at a collider, angular distributions may give an indication of its spin to 
discriminate it against states of spin~0 ({\em e.g.\/}, the sneutrino) and spin~2 (like the Kaluza-Klein graviton 
in extra dimension models). The diagnostics of its 
charges would be of utmost importance as they can hint at the underlying principles.

An interesting class of models is related to $E_6$, a plausible gauge group for unified model building. 
All representations of $E_6$ are free of anomalies so that its $U(1)^\prime$ subgroups 
correspond to $Z^\prime$ candidates.
$Z^\prime$ bosons with the same charges for the SM fermions as in $E_6$ 
also arise within a bottom-up approach~\cite{Erler:2000wu}
when anomaly cancellation is demanded in supersymmetric extensions of the SM 
together with a set of fairly general requirements. 
The breaking chain, $E_6 \to SO(10) \times U(1)_\psi \to SU(5)\times U(1)_\chi \times U(1)_\psi$,
defines a 2-parameter class of models,
\begin{equation}
Z^\prime = \cos\alpha \cos\beta Z_\chi + \sin\alpha \cos\beta Z_Y + \sin\beta Z_\psi,
\end{equation}
where $Y$ denotes hypercharge, and $\alpha \neq 0$ corresponds to the presence of 
a kinetic mixing term $\propto F^{\mu\nu} _YF_{\mu\nu}^\prime$.
Figure~\ref{Zprimes} shows how the combined data from $e^-$ scattering alone 
may cover the entire parameter space for a $Z'$ boson from $E_6$ with a reference mass of 1.2~TeV.
Moreover, the low energy constraints are complementary to other precision constraints on 
$Z'$ bosons~\cite{Erler:2009jh} as well as to collider searches~\cite{Erler:2011ud}.

\begin{figure}
\begin{center}
\includegraphics[height=.48\textheight]{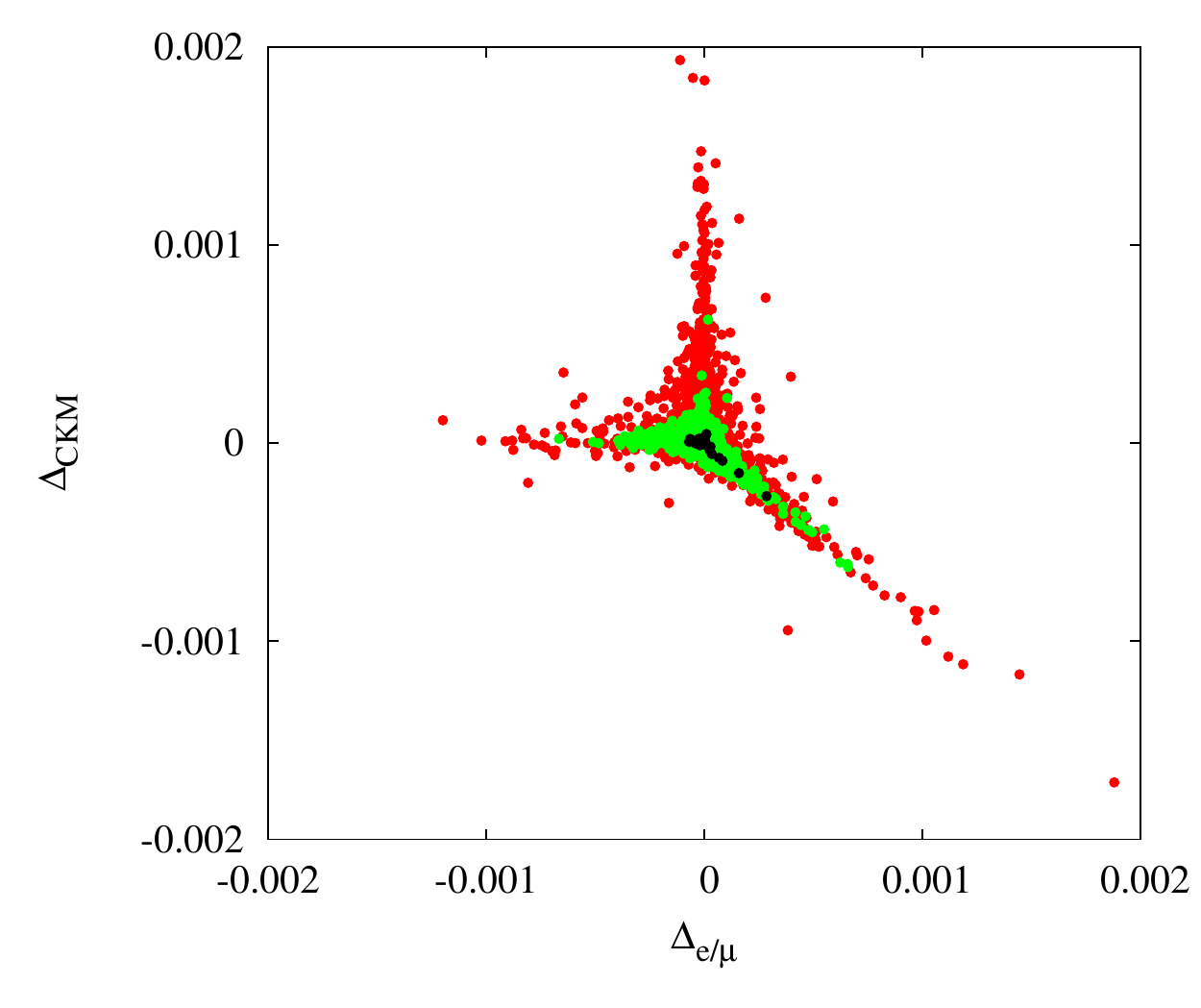}
\end{center}
\caption{Scatter plot~\cite{Bauman:2012fx} of MSSM points satisfying the $\Delta_{\rm CKM}$ and 
$\Delta_{e/\mu}$ constraints.  
Points satisfying in addition the EW precision data including (excluding) LHC bounds are shown in
black (green). 
\label{SUSYCC}}
\end{figure}

\subsection{Charged current observables}

The observables discussed so far are mostly related to the weak NC, but
tests of CC universality can also provide information on new physics.
Deviations from first row unitarity of the CKM matrix are subject to the constraint,
$\Delta_{\rm CKM} \equiv |V_{ud}|^2 + |V_{us}|^2 + |V_{ub}|^2 - 1 = - 0.0001 \pm 0.0006$~\cite{Towner:2010zz}
while those from lepton universality in $\pi^+ \to \ell^+ \nu_\ell (\gamma)$ decays
($\ell = e,\mu$) are constrained by~\cite{Cirigliano:2007ga},
\begin{equation}
\Delta_{e/\mu} \equiv {\Delta R_{e/\mu}\over R_{e/\mu}} \equiv {R_{e/\mu}^{\rm exp}-R_{e/\mu}^{\rm SM}\over R_{e/\mu}^{\rm SM} } 
= - 0.0034 \pm 0.0030 \pm 0.0001,
\end{equation}
and one finds for the minimal supersymmetric standard model (MSSM) the results in Figure~\ref{SUSYCC}.
$\Delta_{\rm CKM}$ is enhanced when there is a large difference between the masses 
of the first generation squarks and the second generation sleptons.
Similarly, $\Delta_{e/\mu}$ is enhanced when the first and
second generation slepton masses are significantly split.

\section{Conclusions}

Precision tests have reached per-mille and sub per-mille accuracy in derived quantities. 
The data are in very good agreement with the SM 
with the only tantalizing deviation sitting in $a_\mu$. 
When combined with the absence of any observation challenging the SM at the LHC, 
this provides tight constraints on new physics and it becomes increasingly likely 
that its energy scale is separated from the SM by at least a little hierarchy. 

\ack
It is a pleasure to thank the organizers of PASCOS 2012 for the invitation to a very enjoyable symposium.
I would also like to thank Sky Bauman, Leo Bellantoni, Jonathan Heckman, Paul Langacker, Shoaib Munir,
Michael Ramsey-Musolf and Eduardo Rojas for collaboration on some of the covered topics.
This work was supported by the CONACyT projects 82291--F and 15 1234.


\section*{References}


\begin{thebibliography}{99}

\bibitem{Glashow:1961tr} 
Glashow S L 1961 {\em Nucl.\ Phys.} {\bf 22} 579

\bibitem{Weinberg:1967tq} 
Weinberg S 1967 {\em Phys.\ Rev.\ Lett.}  {\bf 19} 1264

\bibitem{Prescott:1978tm} 
Prescott C Y {\it et al.} 1978 {\em Phys.\ Lett.} B {\bf 77} 347

\bibitem{Aaltonen:2012ra} 
Aaltonen T {\it et al.} (CDF and D\O\ Collaborations) 2012 {\em Preprint} arXiv:1207.1069 [hep-ex]

\bibitem{TevatronElectroweakWorkingGroup:2012gb} 
Tevatron EW Working Group (CDF and D\O\ Collaborations) 2012 {\em Preprint} arXiv:1204.0042 [hep-ex]

\bibitem{Bennett:2006fi} 
Bennett G W {\it et al.} (Muon g-2 Collaboration) 2006 {\em Phys.\ Rev.} D {\bf 73} 072003
  
\bibitem{Webber:2010zf} 
Webber D M {\it et al.} (MuLan Collaboration) 2011 {\em Phys.\ Rev.\ Lett.} {\bf 106} 041803
  
\bibitem{vanRitbergen:1999fi} 
van Ritbergen T and Stuart R G 2000 {\em Nucl.\ Phys.} B {\bf 564} 343
  
\bibitem{Sirlin:1980nh} 
Sirlin A 1980 {\em Phys.\ Rev.} D {\bf 22} 971
  
\bibitem{PDG2012}  
Erler J and Langacker P 2012 Electroweak Model and Constraints on New Physics
{\em Reference} \cite{Beringer:1900zz}
  
\bibitem{Beringer:1900zz} 
Beringer J {\it et al.} (Particle Data Group) 2012 {\em Phys.\ Rev.} D {\bf 86} 010001
  
\bibitem{Baikov:2008jh} 
Baikov P A, Chetyrkin K G and K\"uhn J H 2008 {\em Phys.\ Rev.\ Lett.} {\bf 101} 012002

\bibitem{Beneke:2008ad} 
Beneke M and Jamin M 2008 {\em J. High Energy Phys.} JHEP09(2008)044

\bibitem{Le Diberder:1992te} 
Le Diberder F and Pich A 1992 {\em Phys.\ Lett.} B {\bf 286} 147

\bibitem{Davier:2008sk} 
Davier M {\em et al.} 2008 {\em Eur.\ Phys.\ J.} C {\bf 56} 305

\bibitem{Boito:2012cr} 
Boito D {\em et al.} 2012 {\em Phys.\ Rev.} D {\bf 85} 093015

\bibitem{Davier:2010nc} 
Davier M, H\"ocker A, Malaescu B and Zhang Z 2011 {\em Eur.\ Phys.\ J.} C {\bf 71} 1515

\bibitem{Arbuzov:1998te} 
Arbuzov A B, Kuraev E A, Merenkov N P and Trentadue L 1998
{\em J. High Energy Phys.} JHEP12(1998)009

\bibitem{Prades:2009tw} 
Prades J, de Rafael E and Vainshtein A 2009 {\em Preprint} arXiv:0901.0306 [hep-ph]
  
\bibitem{Erler:2006vu} 
Erler J and Toledo G 2006 {\em Phys.\ Rev.\ Lett.} {\bf 97} 161801
  
\bibitem{Ellis:1982by} 
Ellis J R, Hagelin J S and Nanopoulos D V 1982 {\em Phys.\ Lett.} B {\bf 116} 283
  
\bibitem{Alcaraz:2006mx} 
Alcaraz J {\it et al.} (ALEPH, DELPHI, L3, OPAL and LEP EW Working Group) 
{\em Preprint} hep-ex/0612034

\bibitem{Zeller:2001hh} 
Zeller G P {\it et al.} (NuTeV Collaboration) 2002 {\em Phys.\ Rev.\ Lett.} {\bf 88} 091802

\bibitem{Gluck:2005xh} 
Gl\"uck M, Jimenez-Delgado P and Reya E 2005 {\em Phys.\ Rev.\ Lett.} {\bf 95} 022002

\bibitem{Sather:1991je} 
Sather E 1992 {\em Phys.\ Lett.} B {\bf 274} 433
  
\nonum
Rodionov E N, Thomas A W and Londergan J T 1994 {\em Mod.\ Phys.\ Lett.} A {\bf 9} 1799
  
\nonum
Martin A D, Roberts R G, Stirling W J and Thorne R S 2004 {\em Eur.\ Phys.\ J.} C {\bf 35} 325
  
\bibitem{Cloet:2009qs} 
Clo\"et I C, Bentz W and Thomas A W 2009 {\em Phys.\ Rev.\ Lett.} {\bf 102} 252301
  
\bibitem{Arbuzov:2004zr} 
Arbuzov A B, Bardin D Y and Kalinovskaya L V 2005 {\em J. High Energy Phys.} JHEP06(2005)078
  
\nonum
Park K, Baur U and Wackeroth D 2009 {\em Preprint} arXiv:0910.5013 [hep-ph]
  
\nonum
Diener K P O, Dittmaier S and Hollik W 2004 {\em Phys.\ Rev.} D {\bf 69} 073005
  
\bibitem{Dobrescu:2003ta} 
Dobrescu B A and Ellis R K 2004 {\em Phys.\ Rev.} D {\bf 69} 114014
  
\bibitem{Diener:2005me} 
Diener K P O, Dittmaier S and Hollik W 2005 {\em Phys.\ Rev.} D {\bf 72} 093002
    
\bibitem{Armstrong:2012ps} 
Armstrong D S {\em et al.} (Qweak Collaboration) 2007 {\em Preprint} arXiv:1202.1255 [physics.ins-det]
  
\bibitem{Gorchtein:2008px} 
Gorshteyn M and Horowitz C J 2009 {\em Phys.\ Rev.\ Lett.} {\bf 102}, 091806
  
\bibitem{Blunden:2011rd} 
Blunden P G, Melnitchouk W and Thomas A W 2011 {\em Phys.\ Rev.\ Lett.} {\bf 107}, 081801

\bibitem{Mammei:2012ph} 
Mammei J (for the MOLLER Collaboration) 2012 {\em Preprint} arXiv:1208.1260 [hep-ex]

\bibitem{Dudek:2012vr} 
Dudek J {\em et al.} 2012 {\em Preprint} arXiv:1208.1244 [hep-ex]
  
\bibitem{Anthony:2005pm} 
Anthony P L {\it et al.} (SLAC--E158 Collaboration) 2005 {\em Phys.\ Rev.\ Lett.} {\bf 95} 081601

\bibitem{Subedi:2011zz} 
Subedi R R {\em et al.} 2011 {\em AIP Conf.\ Proc.} {\bf 1374}, 602

\bibitem{Erler:2004in}
Erler J and Ramsey-Musolf M J 2005 {\em Phys.\ Rev.} D {\bf 72} 073003

\bibitem{:2012an} 
Aad G {\it et al.} (ATLAS Collaboration) 2012 {\em Phys.\ Rev.} D {\bf 86} 032003
  
\bibitem{Chatrchyan:2012tx} 
Chatrchyan S {\it et al.} (CMS Collaboration) 2012 {\em Phys.\ Lett.} B {\bf 710} 26

\bibitem{Wood:1997zq} 
Wood C S {\em et al.} 1997 {\em Science} {\bf 275} 1759
  
\bibitem{Dzuba:2012kx} 
Dzuba V A, Berengut J C, Flambaum V V and Roberts B 2012 
{\em Preprint} arXiv:1207.5864 [hep-ph]

\bibitem{Porsev:2010de} 
Porsev S G, Beloy K and Derevianko A 2010 {\em Phys.\ Rev.} D {\bf 82} 036008
  
\bibitem{Barate:2003sz} 
ALEPH, DELPHI, L3, OPAL and LEP Working Group for Higgs Boson Searches 2003 
{\em Phys.\ Lett.} B {\bf 565} 61
  
\bibitem{TEVNPH:2012ab} 
CDF, D\O\ and Tevatron New Phenomena and Higgs Working Group 2012
{\em Preprint} arXiv:1203.3774 [hep-ex]
  
\bibitem{Erler:2012uu} 
Erler J 2012 {\em Preprint} arXiv:1201.0695 [hep-ph]

\bibitem{Aad:2012gk} 
Aad G {\it et al.} (ATLAS Collaboration) 2012 {\em Phys.\ Lett.} B {\bf 716} 1

\bibitem{Chatrchyan:2012gu} 
Chatrchyan S {\it et al.} (CMS Collaboration) 2012 {\em Phys.\ Lett.} B {\bf 716} 30

\bibitem{CDFandD0:2012zzl} 
CDF, D\O\ and Tevatron New Physics and Higgs Working Group 2012 
{\em Preprint} arXiv:1207.0449 [hep-ex]

\bibitem{Lancaster:2011wr} 
CDF, D\O\ and Tevatron Electroweak Working Group 2011 {\em Preprint} arXiv:1107.5255 [hep-ex]

\bibitem{LHCmt} ATLAS and CMS Collaborations 2012 \\
{\tt https://twiki.cern.ch/twiki/bin/view/CMSPublic/PhysicsResultsTOP12001}

\bibitem{Chetyrkin:1999qi} 
Chetyrkin K G and Steinhauser M 2000 {\em Nucl.\ Phys.} B {\bf 573} 617

\bibitem{Skands:2007zg} 
Skands P Z and Wicke D 2007 {\em Eur.\ Phys.\ J.} C {\bf 52} 133

\bibitem{Fleming:2007qr} 
Fleming S, Hoang A H, Mantry S and Stewart I W 2008 {\em Phys.\ Rev.} D {\bf 77} 074010

\bibitem{Langenfeld:2009wd} 
Langenfeld U, Moch S and Uwer P 2009 {\em Phys. Rev.} D {\bf 80} 054009

\bibitem{Erler:2010sk} 
Erler J and Langacker P 2010 {\em Phys.\ Rev.\ Lett.} {\bf 105} 031801

\bibitem{Kuflik:2012ai} 
Kuflik E, Nir Y and Volansky T 2012 {\em Preprint} arXiv:1204.1975 [hep-ph]

\bibitem{Eberhardt:2012gv} 
Eberhardt O {\em et al.} 2012 {\em Preprint} arXiv:1209.1101 [hep-ph]

\bibitem{Dighe:2012dz} 
Dighe A, Ghosh D, Godbole R M and Prasath A 2012 {\em Phys.\ Rev.} D {\bf 85} 114035
  
\bibitem{Buchkremer:2012yy} 
Buchkremer M, G\'erard J M and Maltoni F 2012 {\em J. High Energy Phys.} JHEP06(2012)135

\bibitem{Bellantoni:2012ag} 
Bellantoni L, Erler J, Heckman J J and Ramirez-Homs E 2012
{\em Preprint} arXiv:1205.5580 [hep-ph]
  
\bibitem{Peskin:1991sw} 
Peskin M E and Takeuchi T 1992 {\em Phys.\ Rev.} D {\bf 46} 381

\bibitem{Jungmann:2008} 
Wansbeek L W {\em et al.} 2008 {\em Phys.\ Rev.} A {\bf 78} 050501(R)

\bibitem{Langacker:2008yv} 
Langacker P 2009 {\em Rev.\ Mod.\ Phys.}  {\bf 81} 1199

\bibitem{Erler:2000wu} 
Erler J 2000 {\em Nucl.\ Phys.} B {\bf 586} 73
  
\bibitem{Erler:2009jh} 
Erler J, Langacker P, Munir S and Rojas E 2009 {\em J. High Energy Phys.}  JHEP08(2009)017

\bibitem{Erler:2011ud} 
Erler J, Langacker P, Munir S and Rojas E 2011 {\em J. High Energy Phys.} JHEP11(2011)076

\bibitem{Towner:2010zz} 
Towner I S and Hardy J C 2010 {\em Rept.\ Prog.\ Phys.} {\bf 73} 046301

\bibitem{Cirigliano:2007ga} 
Cirigliano V and Rosell I 2007 {\em J. High Energy Phys.} JHEP10(2007)005

\bibitem{Bauman:2012fx} 
Bauman S, Erler J and Ramsey-Musolf M J 2012 {\em Preprint} arXiv:1204.0035 [hep-ph]

\end{thebibliography}
\end{document}